\documentclass[%
twocolumn,
preprintnumbers,
superscriptaddress,
nofootinbib,
reprint,
showkeys,
showpacs,
amsmath,amssymb,
aps,
]{revtex4-1}

\usepackage{graphicx}
\usepackage{dcolumn}
\usepackage{bm}
\usepackage{amsfonts,amsbsy}
\usepackage[nodots]{numcompress}
\usepackage{aas_macros}
\usepackage{mathrsfs}
\usepackage{url}
\usepackage{multirow}
\usepackage{xcolor}
\usepackage{color}
\usepackage{ulem}
\usepackage{enumerate}
\usepackage{textcmds}
\usepackage{mathtools}
\usepackage{booktabs}
\usepackage{tabularx}
\usepackage{amsmath}
\usepackage[linkcolor=blue,citecolor=blue,backref=page]{hyperref} 

\renewcommand{\vec}[1]{\boldsymbol{#1}}

\newcommand{\dif}{\mathrm{d}}

\newcolumntype{p}{D{,}{\pm}{-1}}

\begin{document}

\preprint{Version \today}

\title{Probing the nature of Galactic microquasars with diffuse gamma rays and neutrinos: low-mass versus high-mass X-ray binaries}

\author{Pei-Pei Zhang}
\affiliation{School of Intelligence and Information Engineering, Tangshan University, Tangshan, 063000 Hebei, China}
\affiliation{State Key Laboratory of Particle Astrophysics, Institute of High Energy Physics, Chinese Academy of Sciences, Beijing 100049, China}
\affiliation{The Key Laboratory of Cosmic Rays (Xizang University), Ministry of Education, Lhasa, Xizang 850000, China}

\author{Wei Liu}
 \email{liuwei@ihep.ac.cn}
\affiliation{State Key Laboratory of Particle Astrophysics, Institute of High Energy Physics, Chinese Academy of Sciences, Beijing 100049, China}
\affiliation{TIANFU Cosmic Ray Research Center, Chengdu, Sichuan,  China}

\author{Zong-Kang Zeng}
 \email{zengzk@ihep.ac.cn}
 \affiliation{State Key Laboratory of Particle Astrophysics, Institute of High Energy Physics, Chinese Academy of Sciences, Beijing 100049, China}
\affiliation{TIANFU Cosmic Ray Research Center, Chengdu, Sichuan,  China}

 \author{You-Liang Feng}
 \email{fengyouliang@utibet.edu.cn}
\affiliation{The Key Laboratory of Cosmic Rays (Xizang University), Ministry of Education, Lhasa, Xizang 850000, China}

\author{Yi-Qing Guo}
\affiliation{State Key Laboratory of Particle Astrophysics, Institute of High Energy Physics, Chinese Academy of Sciences, Beijing 100049, China}
\affiliation{TIANFU Cosmic Ray Research Center, Chengdu, Sichuan,  China}
\affiliation{University of Chinese Academy of Sciences, Shijingshan District Beijing 100049, China}
\date{\today}

\begin{abstract}
The observations of sub-PeV gamma-ray emission have provided strong evidence for a Galactic origin of the cosmic-ray “knee” region. Meanwhile, mounting observations have ruled out most supernova remnants as sources of PeV cosmic rays, i.e., the so-called PeVatrons. Recently, a new type of PeVatron candidate—microquasar—has received much attention. Studies indicate that microquasars could be the dominant contributor to PeV cosmic rays. However, the spatial distribution of microquasars is not known. They could belong to either LMXBs or HMXBs, depending on the mass of the optical companion star. In this work, we study which type could be the dominant by means of very high energy gamma rays and neutrinos. We find that the diffuse gamma-ray distributions along the Galactic longitude are quite different due to the spatial distributions of these two kinds of sources. If the very high energy gamma rays are dominated by the distribution of LMXBs, the gamma-ray distribution would be more concentrated toward the direction of the Galactic center. Furthermore, we also find that HMXB distribution cannot account for the latest observations of Galactic diffuse neutrinos. Higher-precision measurements of the Galactic diffuse $\gamma$-ray and neutrino fluxes with LHAASO, HAWC, IceCube, and KM3NeT will allow the spatial distribution of the microquasar population to be determined and the origin of the knee to be tested.

\end{abstract}

\keywords{}

\maketitle

\section{Introduction}
\label{sec:intro}

The maximum energy that Galactic cosmic rays (CRs) can reach, and the sources responsible for it, remain open questions in high-energy astrophysics. The CR energy spectrum approximately follows a power-law distribution but shows several pronounced features that may indicate the acceleration limits of Galactic sources: a downward bend at $\sim 3$ PeV  (the ``knee"), where the spectral index $\gamma$ changes from $-2.7$ to $-3.1$; a second steepening at $E \sim 0.1$ EeV (the ``second knee"), with $\gamma$ changing to $-3.4$; a pronounced flattening at $E \sim 5$ EeV (the ``ankle"), with $\gamma$ changing to $-2.7$; and a cutoff at around $50$ EeV \cite{2019PhR...801....1A}. It is commonly believed that the transition from galactic to extra-galactic CRs occurs between ``knee" and ``ankle". However, due to the diffusive propagation of CRs, it is difficult to draw a clear-cut conclusion.

Extremely high-energy electromagnetic radiation, arising as a secondary product of CR acceleration and propagation, may serve as indirect evidence for the presence of PeV CRs in the Galaxy. The H.E.S.S. Collaboration reported the first robust detection of a Galactic PeV source (or ``PeVatrons") within the central $10$ pc of the Galaxy, suggesting Sagittarius A$^\ast$ as the plausible accelerator \citep{2016Natur.531..476H}. Observations of sub-PeV diffuse gamma-ray emission from the Galactic plane provided strong evidence for the existence of PeVatrons in the Galaxy \citep{2021PhRvL.126n1101A}. LHAASO-KM2A further extended the measurements of the diffuse emission from the inner and outer Galactic plane up to $1$ PeV with high significance \citep{2023PhRvL.131o1001C}. Moreover, LHAASO detected PeV gamma-ray emission from the Crab Nebula \citep{2021Sci...373..425L} and reported $12$ ultrahigh-energy gamma-ray sources with energies up to $1.4$ PeV \citep{2021Natur.594...33C}.

Supernova remnants (SNRs) have long been regarded as the primary acceleration sources of Galactic CRs \citep{1934CoMtW...3...79B}, and the non-thermal emission from SNRs substantiates that they are efficient CR accelerators. According to the theory of diffusive shock acceleration (DSA), charged particles are accelerated near strong shock fronts and develop a power-law rigidity spectrum, ${\cal R}^{-\nu}$. The maximum attainable energy of a particle can be estimated using so-called Hillas criterion \citep{1984ARA&A..22..425H, 2005JPhG...31R..95H, 2019IJMPD..2830022G, 2024NatAs...8..425D},
\begin{align}
E_{\rm max}^{\rm th} &= Z \left( \frac{e}{c} \right) {\cal L} ~ U~ {\cal B} \nonumber \\
& \sim \dfrac{Z}{3} \left( \dfrac{\cal L}{\rm pc} \right) \left(\frac{U}{1000 ~\rm km/s} \right) \left( \dfrac{\cal B}{\mu \rm G} \right) ~\rm TeV
\end{align}
The conversion of energy into CRs by SNRs peaks at the transition between the ejecta-dominated and Sedov phases. When typical shock parameters are taken into account, this implies that SNRs can hardly accelerate protons up to the knee region unless the magnetic field at the shock is dramatically amplified. Moreover, gamma-ray observations indicate that most known SNRs appear to cut off at $\sim$ TeV energies, suggesting that CRs can only be accelerated up to several tens of TeV. There is thus little evidence that SNRs can act as PeVatrons.

The findings of LHAASO and other experiments provide a rich and invaluable set of candidate sources for investigating the origin of PeV cosmic rays in our Galaxy. Among the ultrahigh-energy gamma-ray sources are pulsar wind nebulae (PWNe), SNR, young massive star clusters, and others \citep{2021Natur.594...33C}. One class of sources, however, has attracted particular attention: Galactic microquasars. Very recently, the LHAASO Collaboration reported gamma-ray observations of $10$ microquasar candidates. Ultrahigh-energy gamma-ray emission was detected from five of them. The morphology of SS433 and its spatial coincidence with a gas cloud strongly support a hadronic origin for its ultrahigh-energy emission. The energy spectrum of V4641 Sgr extends up to $800$ TeV, indicating that it is a super-PeVatron. Its ultrahigh-energy gamma-ray emission has been detected by the HAWC Collaboration \citep{2024Natur.634..557A}. By extrapolating from the PeV proton luminosity of SS433, one finds that the Galactic microquasar population could supply sufficient CR flux to account for the CR ``knee". Recent X-ray studies conducted at the Galactic center indicate the potential existence of hundreds to thousands of black hole X-ray binaries (BHXBs) in the parsec-scale vicinity of the Galactic center \citep{2018Natur.556...70H, 2021ApJ...921..148M}. These XRBs could play a significant role in shaping the CR spectrum and potentially have a dominant influence slightly above the ``knee" region before the transition to extragalactic sources.

A microquasar is an X-ray binary (XRB) system that launches and collimates relativistic jets. X-ray binaries have long been considered accelerators of Galactic cosmic rays (CRs), converting gravitational potential energy from accreted material or the rotational energy of compact objects into the energy of CR particles, thereby accelerating them \citep{2024JHEAp..43...93K, 2020ApJ...889..146S, 1998NewAR..42..579A}. Several authors have investigated the contribution of microquasars or XRBs to Galactic CRs up to the knee \citep{2025arXiv251001369K, 2025PhRvD.112l3015Z, 2026ApJ..1004..187L, 2026ApJ...997..163Y, 2026arXiv260208940Z, 2026MNRAS.546ag080B, 2026A&A...705L...4A}. However, owing to the paucity of samples, the spatial distribution of microquasars remains poorly constrained. In our previous work, to investigate the contribution of microquasars to the CR knee, we studied the spatial distribution of XRBs as a whole \citep{2026ApJ...997..163Y}.

Depending on the mass of the optical companion, XRBs can be categorized into two classes. One is called Low-mass X-ray binaries (LMXBs) with donor stars less
massive than $\sim 1.5 M_\odot$, and high-mass X-ray binaries (HMXBs) with donor masses exceeding $\sim 10 M_\odot$. The companion of LMXBs can be a white dwarf or a late-type main-sequence star, while for HMXBs, it is typically an O or B star. For the companion masses between $\sim 1.5 M_\odot$ and $\sim 10 M_\odot$, the systems are called intermediate-mass X-ray binaries (IMXBs). As we know, there are few IMXBs discovered in the Galaxy. The observed properties of XRBs are largely defined by the mass-transfer mechanism powering the accretion. In LMXBs, mass exchange occurs through Roche-lobe overflow (RLOF) of the companion, while in HMXBs, the compact star is likely to be fed by the stellar wind or beginning RLOF from the companion.

Owing to their different origins, HMXBs and LMXBs trace different stellar populations and thus exhibit different spatial distributions. In the Galaxy, HMXBs are strongly concentrated toward the Galactic plane and show an association with spiral structure. LMXBs have a stronger concentration toward the Galactic bulge and are also found in the disk and globular clusters.

This paper focuses on distinguishing the spatial distribution of microquasars. Microquasars are assumed to consist primarily of LMXBs or HMXBs and are therefore used to trace the distributions of LMXBs and HMXBs separately. We derive the radial distribution of microquasars by fitting the distributions of LMXBs and HMXBs separately. We further calculate the diffuse gamma-ray energy spectrum and spatial distribution when tracing the spatial distributions of LMXBs and HMXBs.

In addition to diffuse gamma rays, neutrinos are another byproduct of the interaction between cosmic rays (CRs) and the interstellar medium (ISM) during propagation. Recently, the IceCube Collaboration updated the detection of high-energy neutrino emission from the Galactic plane with a significance of $5.7 \sigma$, using a combined analysis of cascades, through-going tracks, and starting tracks. The signal from the Galactic plane is driven largely by an excess of cascade events above $5$ TeV toward the inner Galaxy, within $|l| \leqslant 20^\circ$ and $|b| \leqslant 15^\circ$. We also calculate the separate contributions to the Galactic neutrino flux. We find that when tracing the distribution of HMXBs, Galactic neutrinos are insufficient to account for the latest IceCube observation of the Galactic neutrino flux at tens of TeV.

The paper is organized as follows: in Section \ref{sec:XRB_distribution}, we introduce sample processing methods to describe the spatial distribution of LMXBs and HMXBs. In Sections \ref{sec:model} and \ref{sec:results}, we describe the model and the results. Finally, in Section \ref{sec:summary}, a concluding discussion and an outlook are given.

\section{The Spatial Distribution of LMXBs and HMXBs} \label{sec:XRB_distribution}

\subsection{Low-mass and high-mass X-Ray binaries}
\label{subsec:X-Ray_binaries}

X-ray binaries (XRBs) are among the brightest objects in the sky in the X-ray regime \citep{verbunt1993origin, 2017hsn..book.1499C}. These systems usually consist of a compact object, which can be either a neutron star (NS) or a black hole (BH), and a companion (donor) star. The systems containing a white dwarf (WD) are commonly referred to as cataclysmic variables (CVs) rather than XRBs. The persistent/transient X-ray emission in XRBs originates from the accretion of matter onto the compact object from the companion star. According to Cooper's analysis \citep{2020MNRAS.493.3212C}, which is based on population synthesis findings from \citep{2020A&A...638A..94O}, it is estimated that there are approximately $5500$ XRBs in the Milky Way.

In fact, XRBs can be further categorized into two classes depending on the mass of the optical companion. The high-mass X-ray binaries HMXBs host a compact object (neutron star or black hole) accreting from a massive companion, typically an O or B star with mass exceeding $\sim 10 M_\odot$. The donor is optically bright, and its strong stellar wind often powers the accretion. Three subclasses are recognized: (i) Be/X-ray binaries (BeHMXBs), where accretion occurs from a circumstellar decretion disc; (ii) supergiant X-ray binaries (sgHMXBs), powered by the dense stellar wind of an O/B supergiant; and (iii) Roche-lobe-filling supergiant \citep{2013AdSpR..52.2132C}. The low-mass X-ray binaries (LMXBs), by contrast, contain a low-mass donor (typically $\sim 1.5 M_\odot$), such as a late-type main-sequence star, a white dwarf \citep{Voge2010}. The optical emission is dominated by reprocessed X-ray photons from the accretion disc rather than by the companion itself. Mass transfer proceeds via Roche-lobe overflow through the inner Lagrangian point. The compact object in both HMXBs and LMXBs could be an accreting neutron star or black hole. Therefore, a microquasar could be either LMXB or HMXB. 

Owing to their different origins, HMXBs and LMXBs trace different stellar populations and thus exhibit different spatial distributions. HMXBs are young Population I objects, associated with ongoing star formation. They are concentrated along the Galactic plane, with a scale height of $\approx 150$ pc, and are frequently found in spiral arms and star-forming regions such as the Cygnus region. They tend to avoid the inner $\sim 1$ kpc of the Galaxy \citep{Voge2010}. LMXBs belong to an older Population II component. Their volume density peaks strongly at the Galactic Bulge, and they show no preference for the Galactic plane, with a scale height of $\sim 3-4$ kpc. A substantial fraction of LMXBs are associated with globular clusters, where the formation efficiency is roughly $100$ times higher than in the field.

\subsection{Spatial Distribution}
\label{sec:SD}
HMXBs and LMXBs essentially trace different stellar populations (Population I and Population II objects), with HMXBs lying along the Galactic plane and LMXBs clustering toward the Galactic bulge and in globular clusters. To describe the spatial distribution of LMXBs and HMXBs in the Galaxy, the latest catalogs ~\citep{2023A&A...677A.134N, 2023A&A...675A.199A} have been adopted. It is a complete census of known Galactic XRBs based on X-ray surveys. Up to now, there are $360$ LMXBs and $173$ HMXBs in the catalogs respectively. Based on this catalogue, we can produce the distribution of the LMXBs and HMXBs in the Galaxy.

The comprehensive selection criteria and data processing technique are described in \citep{2026ApJ...997..163Y}, which has been successfully applied to construct the spatial distribution of XRBs. Below, we briefly introduce the method as applied to LMXBs and HMXBs. To correct for distance-dependent selection effects, we employ a gridding method similar to those used to derive the pulsar distribution \citep{yusifov2004revisiting, xie2024modeling}. Quasi-regular grids on the Galactic plane are constructed by drawing concentric circles with equal radial spacing around both the Sun and the Galactic Center (GC). The radial separation between adjacent circles is $\rm \Delta R = 0.85 \rm~kpc$, ensuring a sufficient number of intervals along the Sun–GC axis. The intersections of these circles define a grid, with each grid cell corresponding to a circular region centered on the intersection points.

The spatial distribution of LMXBs is constructed by taking the weighted average of the observed LMXB surface density (SD). To account for Poisson fluctuations of LMXBs ($\rm N_{ij}$) in each spatial grid cell, the error in the observed SD is calculated as
\begin{equation}
\frac{\sigma_{\rho_{ij}}}{\rho_{ij}} = \frac{\sqrt{N_{ij}}}{N_{ij}},
\end{equation}
where $\rm \sigma_{\rho_{i j}}$ denotes the error in $\rm \rho_{i j}$. The radial distribution of LMXBs is then obtained by taking the weighted average:
\begin{equation}
\rho(R_i) = \frac{\sum_j \rho_{ij}\sigma_{ij}^{-2}}{\sum_j \sigma_{ij}^{-2}} \quad \text{and} \quad \sigma^{-2}(R_i) = \sum_j \sigma_{ij}^{-2}
\label{error}
\end{equation}
where $i$ and $j$ represent the grid indices corresponding to the position($R_{i}$, $r_{j}$), with $R_{i}=i \times 0.85~kpc$ being the distance from the GC, and $r_{j}=j \times 0.85~kpc$ the distance from the Sun. The same procedure is applied to HMXBs.

\begin{figure}[ht!]
\centering
\includegraphics[width=\columnwidth]{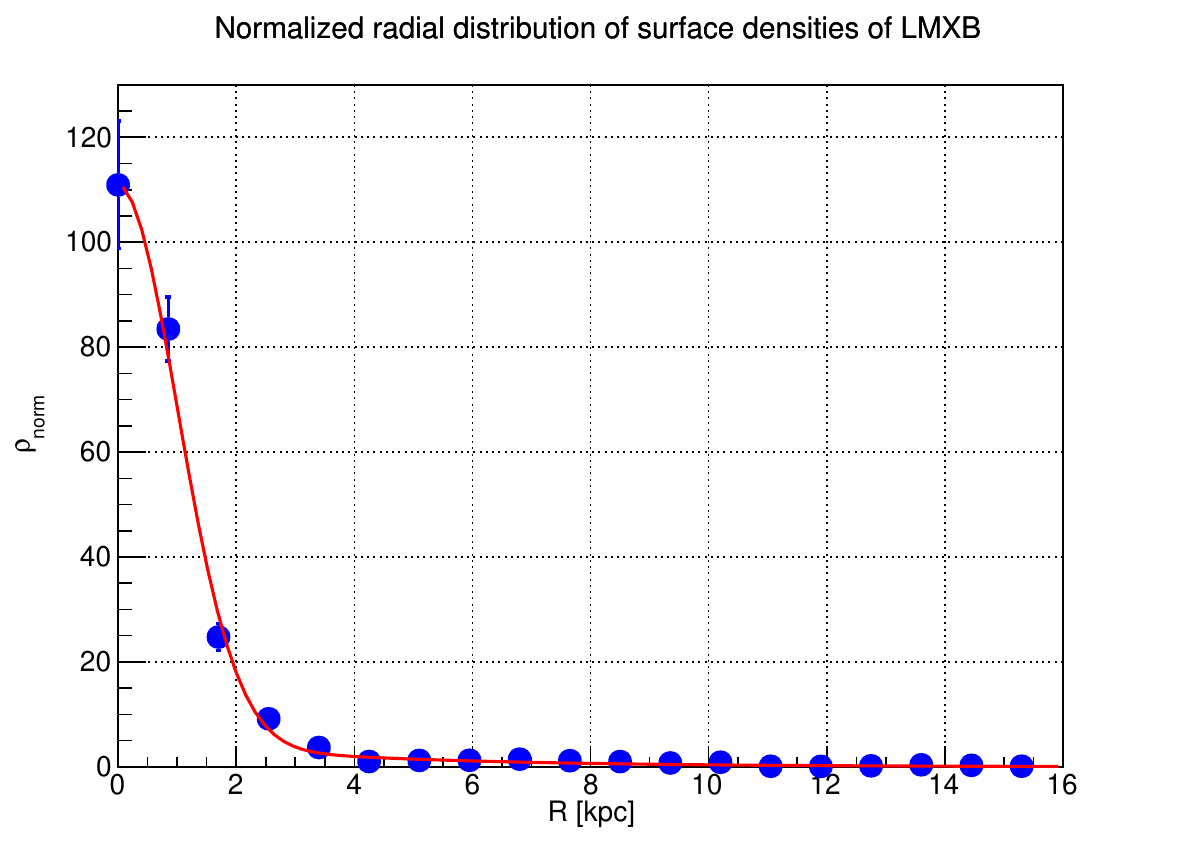}
\includegraphics[width=\columnwidth]{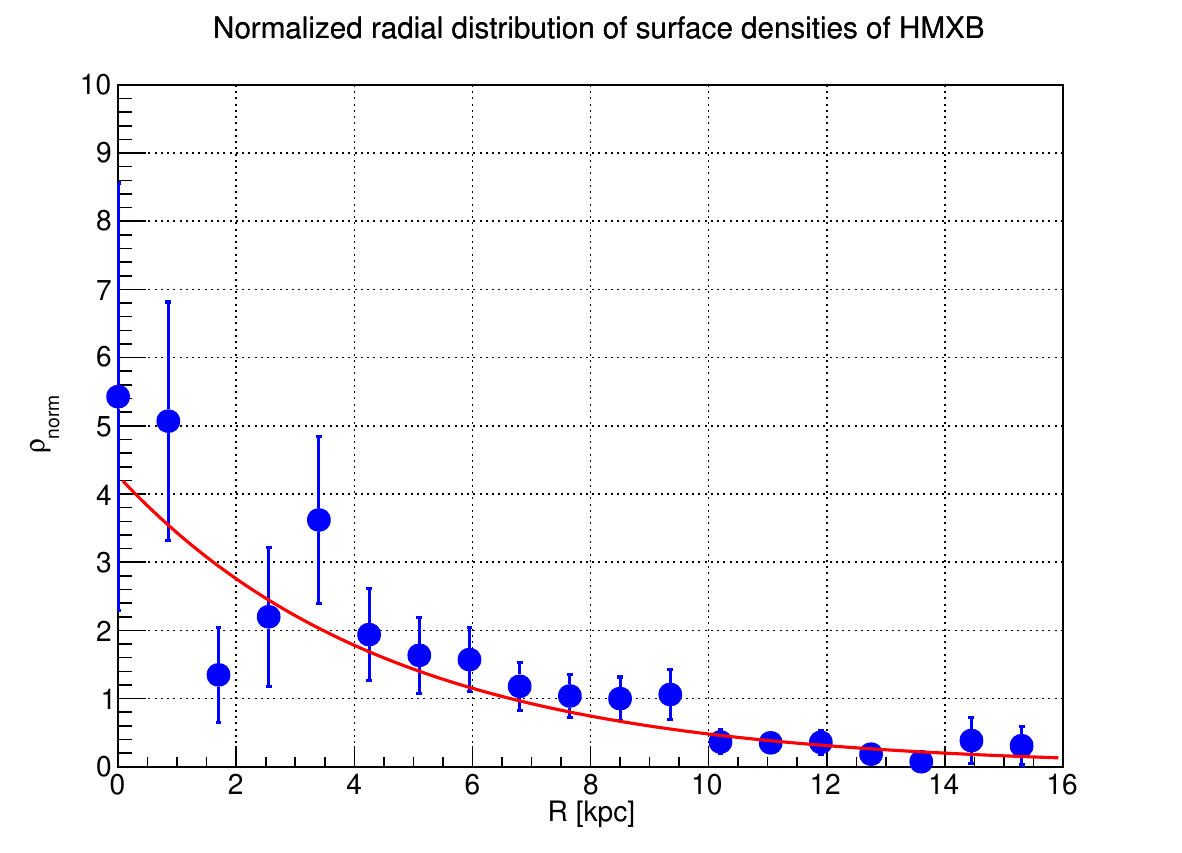}
\caption{Normalized radial distributions of the surface densities of LMXBs (upper panel) and HMXBs (lower panel) derived from the latest XRB catalogues\citep{2023A&A...677A.134N,2023A&A...675A.199A}. The red solid lines show the fits with Eqs.~(\ref{eq:rho_lmxb_fit}) and (\ref{eq:rho_hmxb_fit}). The data points and the fitted curve in the figure have been normalized to the value at the Sun’s location.}
\label{LMXBs_HMXBs_distri}
\end{figure}

The radial distributions of LMXBs and HMXBs are fitted by
\begin{align}
\rho(r)_{\rm LMXB} = A \left[ \exp\left(-a \frac{r}{r_{\odot}} \right) + B \times \exp\left(-b \left(\frac{r}{r_{\odot} } \right)^2\right) \right]
\label{eq:rho_lmxb_fit}
\end{align}
and
\begin{align}
\rho(r)_{\rm HMXB} = A \exp \left(-a \frac{r}{r_{\odot} }\right)
\label{eq:rho_hmxb_fit}
\end{align}
respectively, with $r_\odot \equiv 8.5$ kpc representing the distance from the Galactic center to the solar system. The results are shown in Figure \ref{LMXBs_HMXBs_distri}. For LMXBs, $a = 2.31$, $b =35.7$, $A = 0.79 ~\rm/kpc^{2}$ and $\rm B = 18.3$, while for HMXBs, $A = 0.44$~/kpc$^2$ and $a = 1.86$. The source distribution of LMXBs and HMXBs normalized at the solar position can be written as:
\begin{equation}
\rho(R, z)_{\rm LMXB} = \frac{\exp (-a\frac{r}{r_{\odot}})+B\exp (-b\frac{r^{2}}{r_{\odot}^{2}})}{\exp (-a)+B\exp (-b)}\exp \left(-\frac{|z|}{z_s}\right),
\label{LMXB_distri_norm}
\end{equation}
and
\begin{equation}
\rho(R, z)_{\rm HMXB} = \frac{\exp (-a\frac{r}{r_{\odot}})}{\exp (-a)}\exp \left(-\frac{|z|}{z_s}\right),
\label{HMXB_distri_norm}
\end{equation}
where $\rm z_s = 0.2~ kpc$.

\section{Model Description}
\label{sec:model}

\subsection{Cosmic Ray Transport}
\label{sec:crt}

In this work, the spatially-dependent propagation (SDP) model is adopted to describe cosmic-ray (CR) propagation in the Galaxy. This framework has been widely applied to calculations of CR energy spectra \citep{2012ApJ...752L..13T, 2015PhRvD..92h1301T, 2016PhRvD..94l3007F, 2016ApJ...819...54G, 2018ApJ...869..176L, 2018PhRvD..97f3008G, 2020ChPhC..44h5102T, 2026ApJ..1004..187L}, the diffuse gamma-ray distribution, and large-scale anisotropies \citep{2019JCAP...10..010L, 2019JCAP...12..007Q, 2021FrPhy..1624501Y, 2026ApJ...997..163Y}. For a detailed introduction, we refer the reader to the above references. Here we only give a brief overview.

In the SDP model, the diffusive halo is divided into two parts: the inner and outer halo. Within the inner halo, the diffusion coefficient is much smaller than that in the outer halo, as indicated by the HAWC observations. The shape of the magnetic halo is usually approximated as a cylinder, with the Galactic disk embedded in the middle. The radial boundary of the magnetic halo is set to the Galactic radius, i.e., $R = 20~\mathrm{kpc}$, whereas its half-thickness $z_h$ needs to be evaluated by fitting the CR data \citep{2017PhRvD..95h3007Y}. Both CR sources and the interstellar medium (ISM) are mainly distributed within the Galactic disk, whose width $z_s$ is set to be $\sim 200~\mathrm{pc}$. Aside from the diffusion effect, CR nuclei may also undergo convection, reacceleration, and fragmentation due to collisions with interstellar gas. At lower energies, CR nuclei further lose energy via ionization and Coulomb scattering. The transport equation is generally written as
\begin{equation}
\begin{aligned}
\frac{\partial \psi(\mathbf{r}, p, t)}{\partial t}
=\, & Q(\mathbf{r}, p, t)
+ \nabla \cdot \left( D_{xx} \nabla \psi
- \mathbf{V}_c \psi \right) \\
& + \frac{\partial}{\partial p}
\left[
p^{2} D_{pp}
\frac{\partial}{\partial p}
\left(
\frac{1}{p^{2}} \psi
\right)
\right] \\
& - \frac{\partial}{\partial p}
\left[
\dot{p}\, \psi
- \frac{p}{3}
(\nabla \cdot \mathbf{V}_c)\, \psi
\right] \\
& - \frac{\psi}{\tau_f}
- \frac{\psi}{\tau_r} ,
\end{aligned}
\label{eq:CR_transport}
\end{equation}
where $\psi = dn/dp$ is the CR density per unit particle momentum $p$
at position $\mathbf{r}$, $Q(\mathbf{r}, p, t)$ is the source function,
$D_{xx}$ and $D_{pp}$ are the diffusion coefficients in the spatial and
momentum spaces, respectively, $\mathbf{V}_c$ is the convection velocity,
$\dot{p}$ is the momentum loss rate, and $\tau_f$ and $\tau_r$ denote the fragmentation and radioactive decay timescales. At the boundary of the halo, free escape of CRs is assumed, namely, $\psi(R, z, p) = \psi(r, \pm z_h, p) = 0$.

The spatial diffusion coefficient $D_{xx}$ can be parameterized as
\begin{equation}
D_{xx}(r,z, {\cal R} )= D_{0}F(r,z)\beta^{\eta} \left(\dfrac{\cal R}
{{\cal R}_{0}} \right)^{\delta_{0}F(r,z)},
\label{eq:diffusion}
\end{equation}
\begin{equation}
F(r,z) = \left\{
\begin{array}{ll}
{\dfrac{N_m}{1+f(r,z)}+\left[1-\dfrac{N_{m}}{1+f(r,z)}\right]}\left(\dfrac{{z}}{\xi{z}_{\rm h}} \right)^{n},  &  {{|z|} \leq \xi{z}_{\rm h}} \\
\\
1,  &  { {|z|} > \xi{z}_{\rm h}} \\
\end{array},
\right.
\end{equation}
where $r$ and $z$ are cylindrical coordinates, ${\cal R}$ is the particle's rigidity, $\beta$ is the particle's velocity in units of light speed, $D_0$ and $\eta$ are constants. The half-thickness of the propagation halo is $z_h$, and the
half-thickness of the inner halo is $\xi z_{h}$.

In this work, we adopt the diffusion reacceleration model, with the diffusive-reacceleration coefficient $D_{pp}$, which correlated with $D_{xx}$ via $D_{pp}D_{xx} = \dfrac{4p^{2}v_{A}^{2}}{3\delta(4-\delta^{2})
(4-\delta)}$, where $v_A$ is the Alfv\'en velocity, $p$ is the momentum,
and $\delta$ is the rigidity dependence slope of the diffusion coefficient \citep{1994ApJ...431..705S}. The numerical package DRAGON is used to solve the propagation equation of CRs \citep{2017JCAP...02..015E}.

\subsection{Background Supernova Remnants}
\label{subsec:bkg}
The spatial distribution of background SNRs is approximated as an axisymmetric form, which can be parameterized as
\begin{equation}
f(r, z) = \left(\dfrac{r}{r_\odot} \right)^\alpha \exp \left[-\dfrac{\beta(r-r_\odot)}{r_\odot} \right] \exp \left(-\dfrac{|z|}{z_s} \right) ~,
\label{eq:radial_dis}
\end{equation}
Parameters $\alpha$ and $\beta$ are taken as $1.69$ and $3.33$ in this work \citep{1996A&AS..120C.437C}. The density of the SNR distribution decreases exponentially along the vertical height from the Galactic plane, with $z_{s} = 200$ pc. The injection spectrum of nuclei is assumed to be a broken power-law function of particle rigidity with an exponential cut off:
\begin{equation}
q({\cal R}) \propto \left(\dfrac{\cal R}{ {\cal R}_{\rm 0} }\right)^{-\nu} \exp \left( -\dfrac{\cal R}{{\cal R}_{\rm c}} \right) ~.
\end{equation}

\subsection{Nearby Young Supernova Remnant}
\label{subsec:nySNR}

In this work, a local young SNR is invoked to explain the fine structure of spectral hardening and break-off between $\sim 200$ GeV and $\sim 14$ TeV \citep{2016PhRvL.117o1103A, 2019JCAP...10..010L, 2019JCAP...12..007Q, 2021PhRvD.104j3013F,  2022ApJ...926...41Z, 2022MNRAS.511.6218Z, 2024ApJ...962...43L, 2026ApJ..1004..187L}. The progenitor of Geminga, an SNR, was introduced as the local source. The injection process of the SNR is approximated as a burst-like process. The source injection rate as a function of time and rigidity is assumed to be
\begin{equation}
Q({\cal R}, \vec{r}, t)= q_{0} \left(\dfrac{\cal R}{{\cal R}_0}
\right)^{-\gamma} \exp \left( -\dfrac{\cal R}{{\cal R}_{\rm c}}
\right) \delta(\vec{r} -\vec{r}_0) \delta(t-t_0) ~,
\label{eq:nearby}
\end{equation}
where  ${\cal R}_{\rm c}$ is the cutoff rigidity and $\vec{r}_0$ and $t_0$ are initial position and time of the supernova explosion. The propagated spectrum from Geminga SNR is thus a convolution of the Green's function and the time-dependent injection rate $Q({\cal R}, \vec{r}, t)$
\begin{equation}
\varphi(\vec{r}, {\cal R}, t) = \int \int G(\vec{r}-\vec{r}^\prime, t-t^\prime, {\cal R}) Q({\cal R}, \vec{r}^\prime, t^\prime) \dif \vec{r}^\prime \dif  t^\prime .
\end{equation}

\subsection{X-ray Binaries}
\label{subsec:xrb}
The CR source term of the XRB population is constructed as follows. The spatial distribution of XRBs follows Eqs.~(\ref{LMXB_distri_norm}) and (\ref{HMXB_distri_norm}), derived in Section~\ref{sec:SD} from the latest catalogues. The injection spectrum of CR nuclei from the XRB population is
assumed to be a power law in rigidity with an exponential cutoff,
\begin{equation}
q_{\rm XB}(R) \propto \left(\frac{R}{R_{0}}\right)^{-\gamma_{\rm XB}}
\exp\left(-\frac{R}{R_{c}^{\rm XB}}\right),
\label{eq:xrb_injection}
\end{equation}
motivated by particle acceleration in microquasar jets\citep{2025arXiv251001369K, 2025PhRvD.112l3015Z}. Since the acceleration is predominantly hadronic, only CR nuclei are injected by the XRB component; the XRBs therefore contribute to the diffuse emission only through $\pi^{0}$ decay. 

The cutoff rigidity and the injection index are fixed at $R_{c}^{\rm XB}=5$~PV and $\gamma_{\rm XB}=2.1$ for all scenarios, while the normalization of each scenario is determined by requiring that the total proton spectrum reproduces the measurements (Section~\ref{sec:proton}).
Three scenarios are considered. In the XB\_All scenario, the source distribution is taken from \citep{2026ApJ...997..163Y}.
The resulting injection parameters are summarized in Table~\ref{tab:xrb_para}.

\section{Results}
\label{sec:results}

Following the methodology described in Section~\ref{sec:model}, we compute the proton
spectrum for each of the three XRB scenarios and adjust the normalization of the XRB
component so that the total spectrum reproduces the measured proton spectrum. The
diffuse $\gamma$-ray and neutrino emissions are then derived within the same framework
without any further tuning. We present the proton spectrum, the diffuse $\gamma$-ray
emission, and the diffuse neutrino emission in turn below.

\subsection{The proton spectrum}
\label{sec:proton}

As promising candidates for Galactic PeVatrons, X-ray binaries (microquasars) may play a key role in shaping the CR spectrum around the knee. We therefore adopt a framework in which the proton spectrum at low energies is dominated by the SNR population together with a nearby (local) SNR, while the contribution from XRBs takes over at the highest energies and accounts for the spectrum up to and beyond the knee. Within
this framework, we compute the proton spectrum for three scenarios, corresponding to the entire XRB population (XB\_All), an HMXB-dominated case, and an LMXB-dominated case. The SNR and nearby-source components are common to all three scenarios, whereas the XRB component varies according to the adopted spatial distribution of each population. The propagation parameters of the SDP model ($D_0$, $\delta_0$, $N_m$, $\xi$, $n$, $\eta$, $v_A$, and $z_h$) are taken from \citep{2026ApJ...997..163Y} and the injection parameters of the SNR and nearby-source components are summarized in Table~\ref{tab:para_all}.

\begin{figure}[ht!]
\centering
\includegraphics[width=\columnwidth]{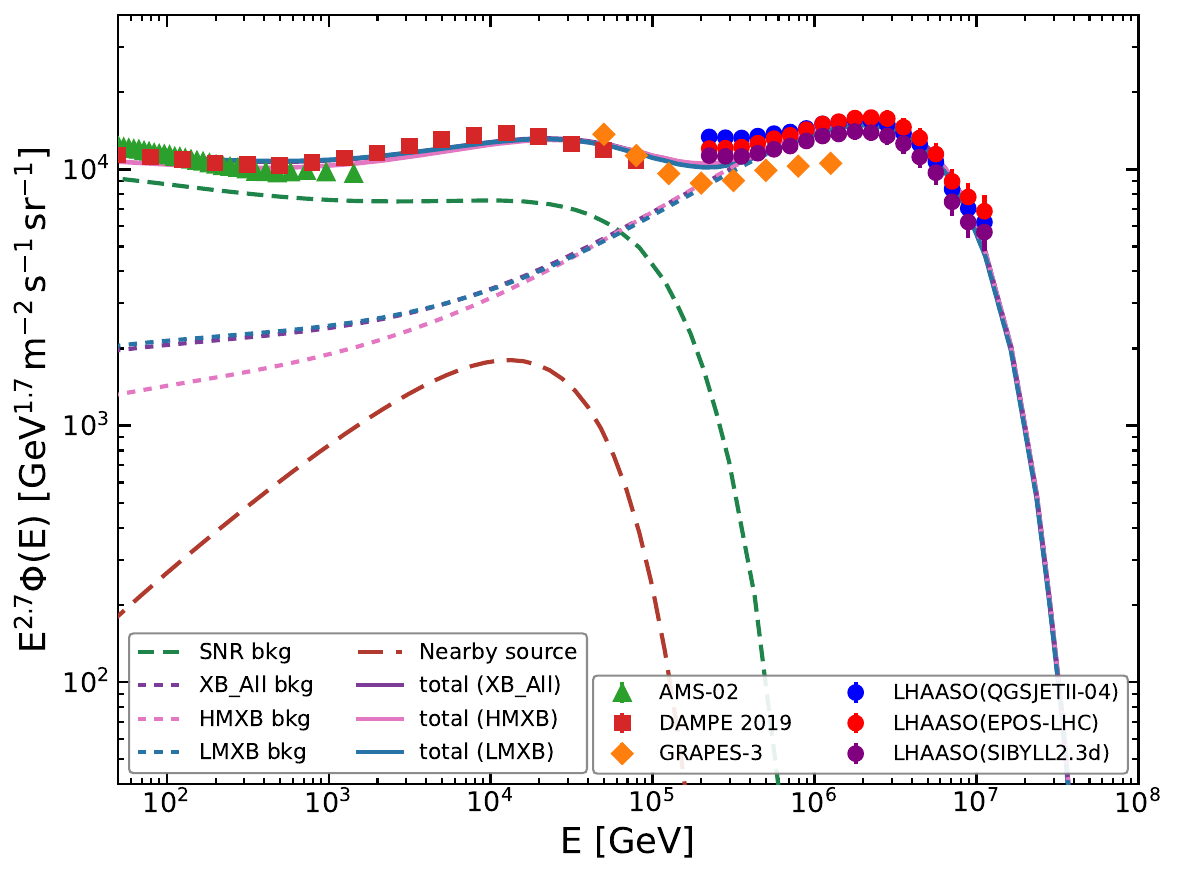}
\caption{The calculated proton spectrum for the XB\_All(purple), HMXB-dominated (pink), and LMXB-dominated (blue) scenarios, compared with observed data from AMS-02\citep{AMS:2015tnn}, DAMPE\citep{DAMPE:2019gys}, GRAPES-3\citep{GRAPES-3:2024mhy} and LHAASO\citep{LHAASO:2025byy}. Dashed lines: X-ray binary contribution; solid lines: total spectrum. Green long-dashed and red dashed lines: SNR and nearby-source contributions, respectively.}
\label{fig_proton_spe}
\end{figure}

Figure~\ref{fig_proton_spe} compares the calculated proton spectra with the measurements of AMS-02, DAMPE, GRAPES-3, and LHAASO. All three scenarios reproduce the observed spectrum well over the entire energy range. So the proton spectrum alone cannot discriminate, the discrimination comes entirely from the spatial morphology. In particular, the agreement extends across the knee region, where the SNR and nearby-source components have
already cut off and the spectrum is dominated by the XRB contribution.

\begin{table*}[t!]
\begin{center}
\begin{tabular}{|ccccccccc|}
\toprule[1.5pt]
& $R_c^{\rm bkg}$ & $\nu$ & Normalization & $q_0$ & $\gamma$  & $R_c$ & $r_0$ & $t_0$\\
& [TeV] &  & $[{\rm GeV}^{-1}{\rm m}^{-2} {\rm sr^{-1}} {\rm s^{-1}}]$ &[GeV$^{-1}$] &   & [TeV] & [kpc] & [yr]\\
& $100$ & $2.3$  & $3.37\times 10^{-2}$  & $2.8\times 10^{52}$ & $2.08$ & $10$ & 0.3 & $3\times 10^5$\\
\hline
\bottomrule[1.5pt]
\end{tabular}
\end{center}
\caption{Parameters of the background CR injection and the nearby source adopted in this work.}
\label{tab:para_all}
\end{table*}

\begin{table}[t!]
\begin{center}
\begin{tabular}{|cc|}
\toprule[1.5pt]
Scenario & Normalization $[{\rm GeV}^{-1}{\rm m}^{-2} {\rm sr^{-1}} {\rm s^{-1}}]$ \\
\hline
XB\_All & $7.9\times 10^{-3}$ \\
HMXB-dominated & $5.5\times 10^{-3}$ \\
LMXB-dominated & $8.2\times 10^{-3}$ \\
\hline
\bottomrule[1.5pt]
\end{tabular}
\end{center}
\caption{Injection parameters of the XRB component for the three scenarios fixed at $\gamma_{XB}$=2.1, $R_c^{XB}$=5 PV, the normalization of
Eq.~(\ref{eq:xrb_injection}), determined by fitting the proton spectrum.}
\label{tab:xrb_para}
\end{table}

\subsection{Diffuse gamma-rays}
\label{sec:diffuse_gamma}

Diffuse gamma-ray emission (DGE) in the Galaxy is expected to be produced by interactions of cosmic rays (CRs) with the interstellar medium (ISM) and the interstellar radiation field (ISRF) during the propagation of CRs in the Milky Way. DGE therefore provide a means to constrain the source distribution and propagation mechanisms of Galactic CRs. The DGE consists mainly of three components: $\pi_0$ decay, from inelastic hadronic interactions between CR nuclei and the ISM; bremsstrahlung, from CREs in the ISM; and inverse Compton scattering (ICS), from the scattering of CREs off the ISRF. 

\begin{figure*}[t!]
\centering
\includegraphics[width=0.48\textwidth]{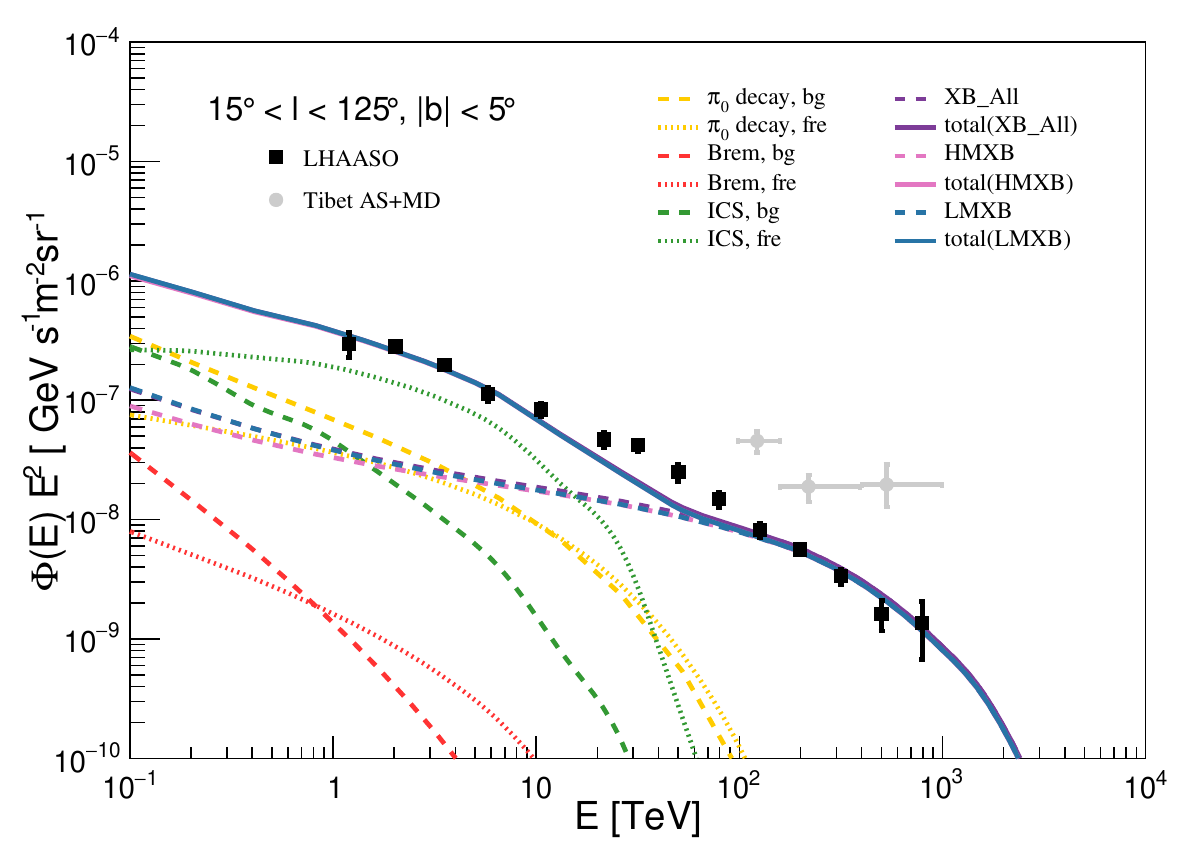}
\hfill
\includegraphics[width=0.48\textwidth]{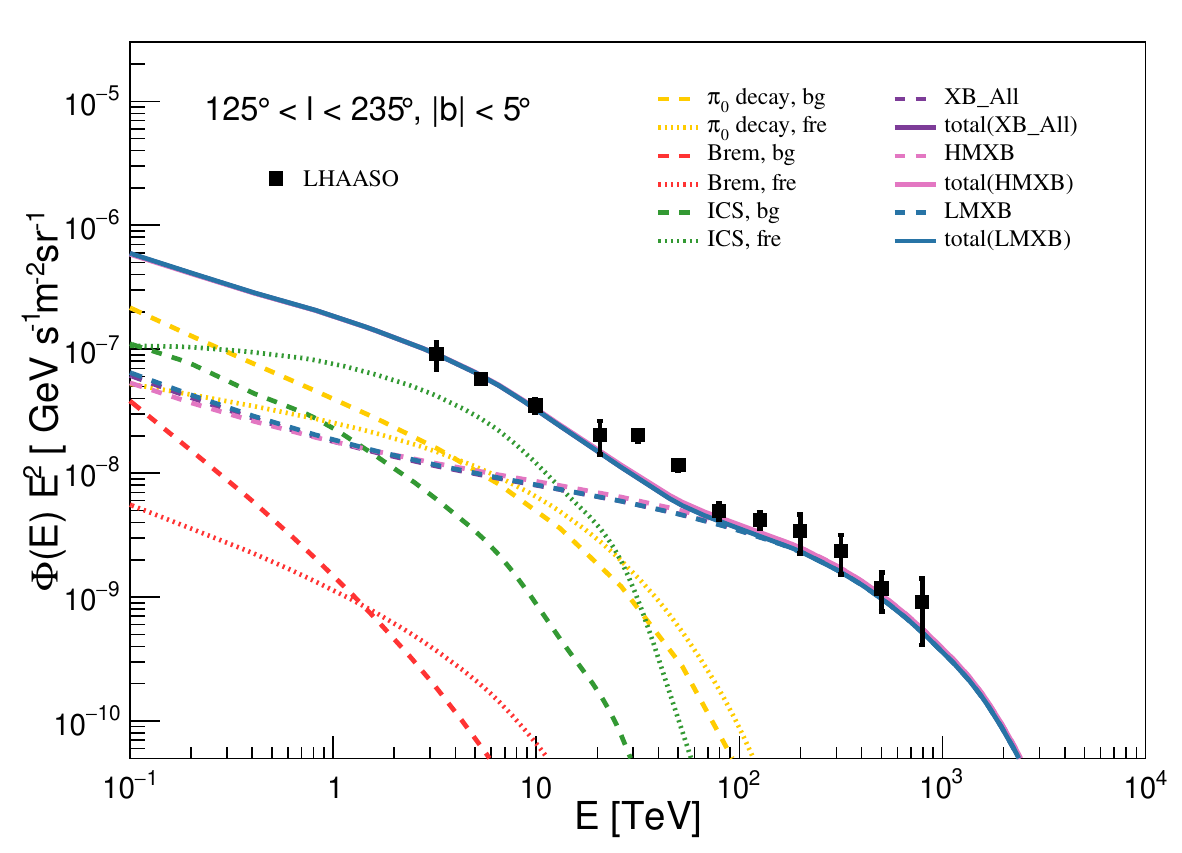}
\caption{Diffuse $\gamma$-ray emission of the inner region ($15^\circ < l < 125^\circ$, left) and the outer region ($125^\circ < l < 235^\circ$, right) of the Galaxy, both with $|b| < 5^\circ$.  In each panel, the total emission for the XB\_All, HMXB-dominated, and LMXB-dominated scenarios is shown as purple, pink, and blue solid lines, respectively, with the corresponding X-ray binary contributions given by the same-colored dashed lines. The SNR contributions from $\pi_0$ decay, bremsstrahlung, and inverse Compton scattering (ICS) are shown as yellow, red, and green lines, respectively, with dotted and dash-dotted styles denoting the background and fresh components. Observational data are taken from LHAASO\citep{2025PhRvL.134h1002C} and Tibet AS+MD\citep{2021PhRvL.126n1101A}.}
\label{fig_diffuse_gamma}
\end{figure*}

Figure~\ref{fig_diffuse_gamma} presents the diffuse $\gamma$-ray emission predicted by our model in the inner ($15^{\circ}<l<125^{\circ}$) and outer ($125^{\circ}<l<235^{\circ}$) regions of the Galactic plane ($|b|<5^{\circ}$), in comparison with the measurements of LHAASO \citep{2025PhRvL.134h1002C} and Tibet AS+MD \citep{2021PhRvL.126n1101A}. In our model, the same masks as those adopted in the LHAASO analysis are applied, ensuring a self-consistent comparison. The predicted flux consists of the SNR component --- $\pi^{0}$ decay, bremsstrahlung, and inverse Compton scattering (ICS) produced by the CRs injected from both the background and the fresh sources --- and the contribution from the XB population under the three scenarios introduced above.

The composition of the diffuse emission varies significantly with energy. Below $\sim 10$~TeV, the total flux is dominated by the SNR component. Among the individual processes, the $\pi^{0}$ decay of the background CRs provides the largest contribution over most of this range, while the ICS emission, including both the background and the fresh components, is comparable to the $\pi^{0}$-decay emission below $\sim 1$~TeV and even becomes the leading component at sub-TeV energies in the outer region; bremsstrahlung remains subdominant at all energies. Above $\sim 10$~TeV, the SNR components decline rapidly following the cutoffs of the source spectra, and the XB contribution takes over, accounting for the bulk of the diffuse flux up to PeV energies. The three XB scenarios result in very similar total fluxes.

The predicted total flux agrees well with the LHAASO measurement from $\sim$TeV up to $\sim$PeV energies in both the inner and outer regions. In particular, the spectral steepening of the diffuse emission above $\sim 100$~TeV revealed by LHAASO is naturally reproduced by the cutoff rigidities of the source spectra (Table~\ref{tab:para_all}), and the inner-to-outer flux ratio of about two, which traces the radial distribution of the CR sources, is also well accounted for.

\begin{figure}[ht!]
\centering
\includegraphics[width=\columnwidth]{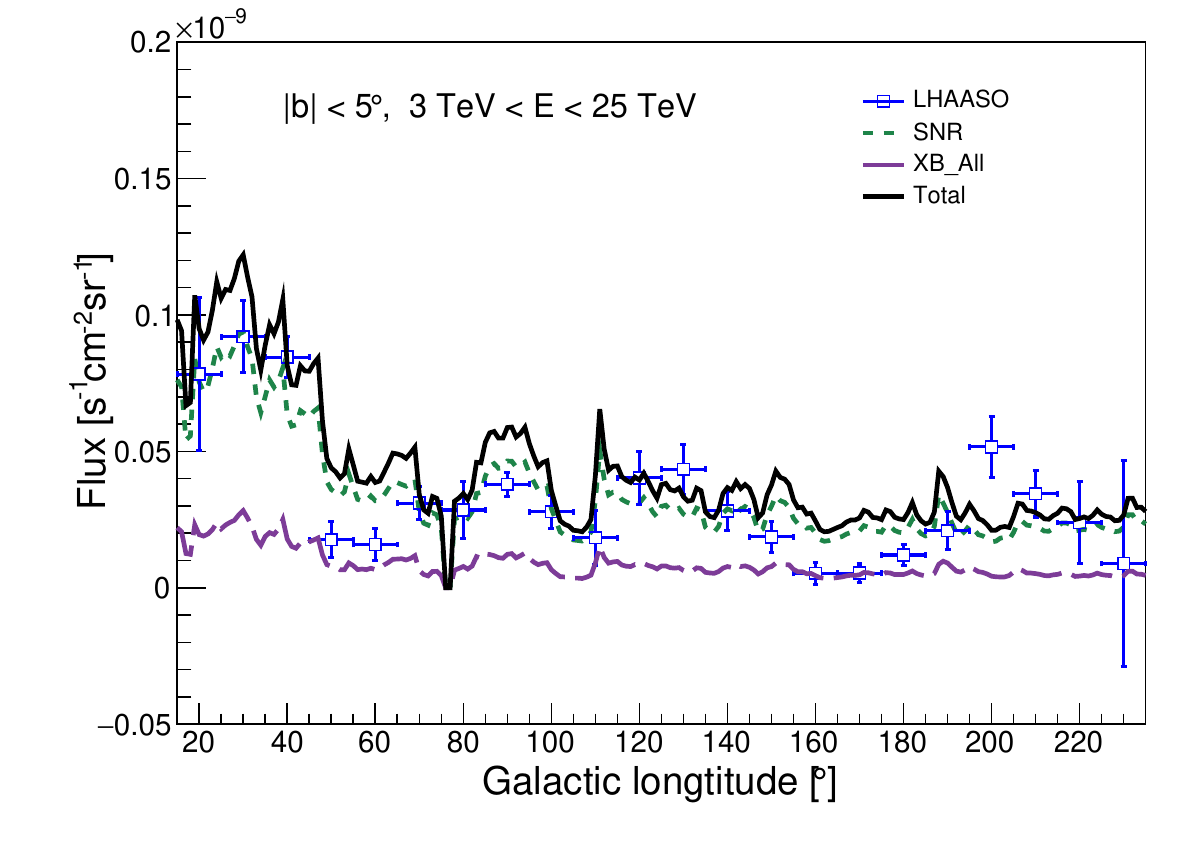}
\includegraphics[width=\columnwidth]{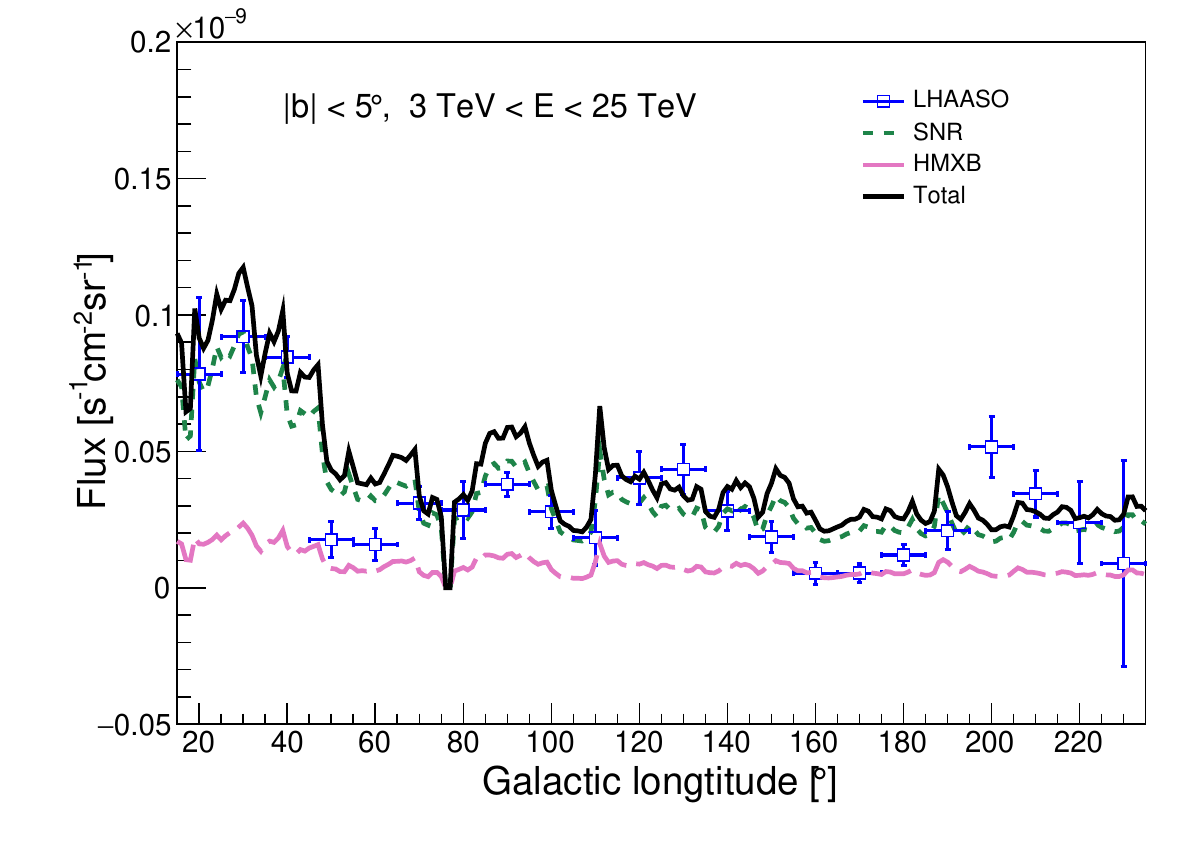}
\includegraphics[width=\columnwidth]{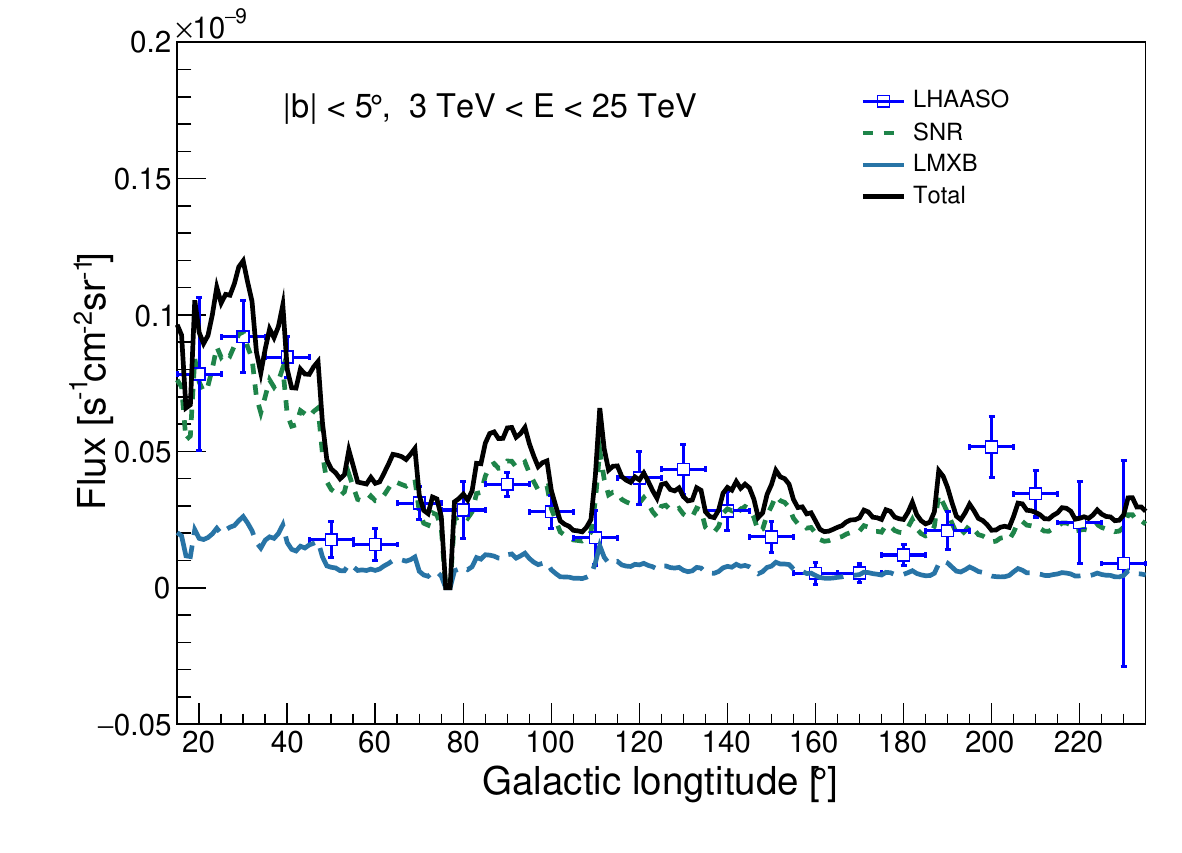}
\caption{Galactic longitude profiles of the diffuse $\gamma$-ray flux integrated over $3\ {\rm TeV}<E<25\ {\rm TeV}$ for $|b|<5^{\circ}$, compared with the LHAASO measurement \citep{2025PhRvL.134h1002C}. The three panels correspond to the XB\_All (top), HMXB-dominated (middle), and LMXB-dominated (bottom) scenarios. In each panel, the green line shows the SNR contribution, the purple, pink, and blue line the XRB contribution, and the black line the total. }
\label{fig_gde}
\end{figure}

\begin{figure}[ht!]
\centering
\includegraphics[width=\columnwidth]{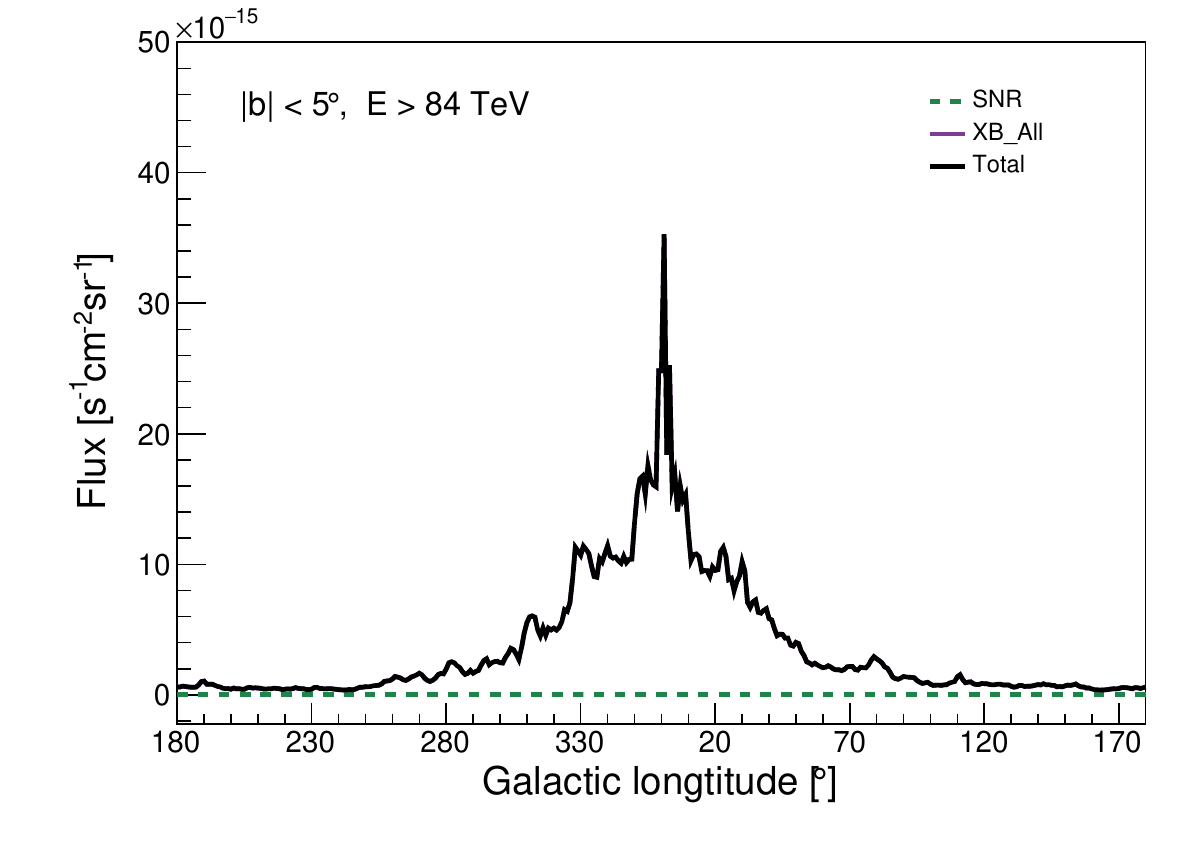}
\includegraphics[width=\columnwidth]{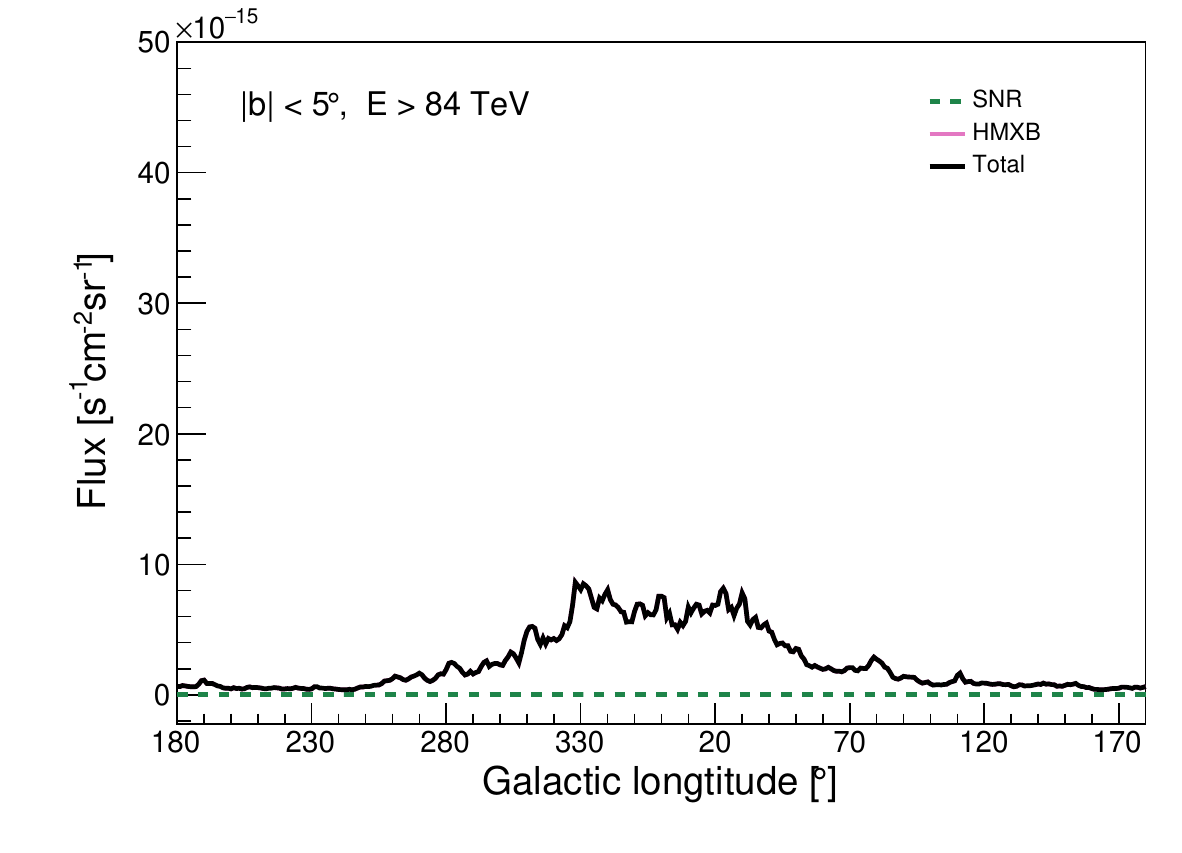}
\includegraphics[width=\columnwidth]{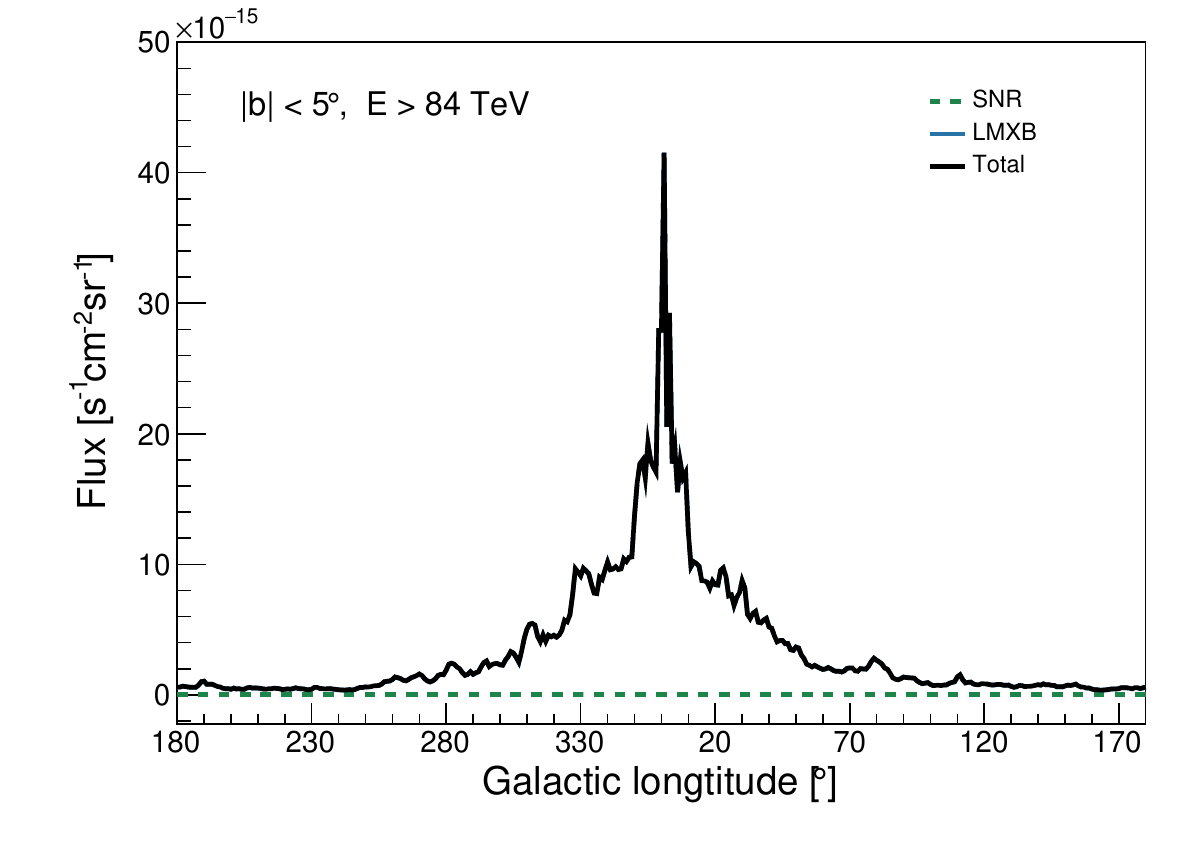}
\caption{Same as Figure~\ref{fig_gde}, but integrated over $E>84$~TeV.}
\label{fig_gde2}
\end{figure}

The different spatial distributions of the two XRB sub-populations leave a direct imprint on the Galactic longitude profiles of the diffuse emission. Figures~\ref{fig_gde} and \ref{fig_gde2} show the predicted profiles for $|b|<5^{\circ}$, integrated over $3\ {\rm TeV}<E<25\ {\rm TeV}$ and $E>84\ {\rm TeV}$, respectively, in comparison with the LHAASO measurements \citep{2025PhRvL.134h1002C}. In the LMXB-dominated scenario the emission is more concentrated toward the Galactic center, reflecting the bulge component of the LMXB distribution, whereas the HMXB-dominated scenario yields a flatter profile that follows
the disk.With the current measurement uncertainties the three scenarios cannot yet be distinguished by the longitude profiles alone, but future higher-statistics observations will be sensitive to this difference.

\subsection{Diffuse neutrinos}

\begin{figure}[ht!]
\centering
\includegraphics[width=\columnwidth]{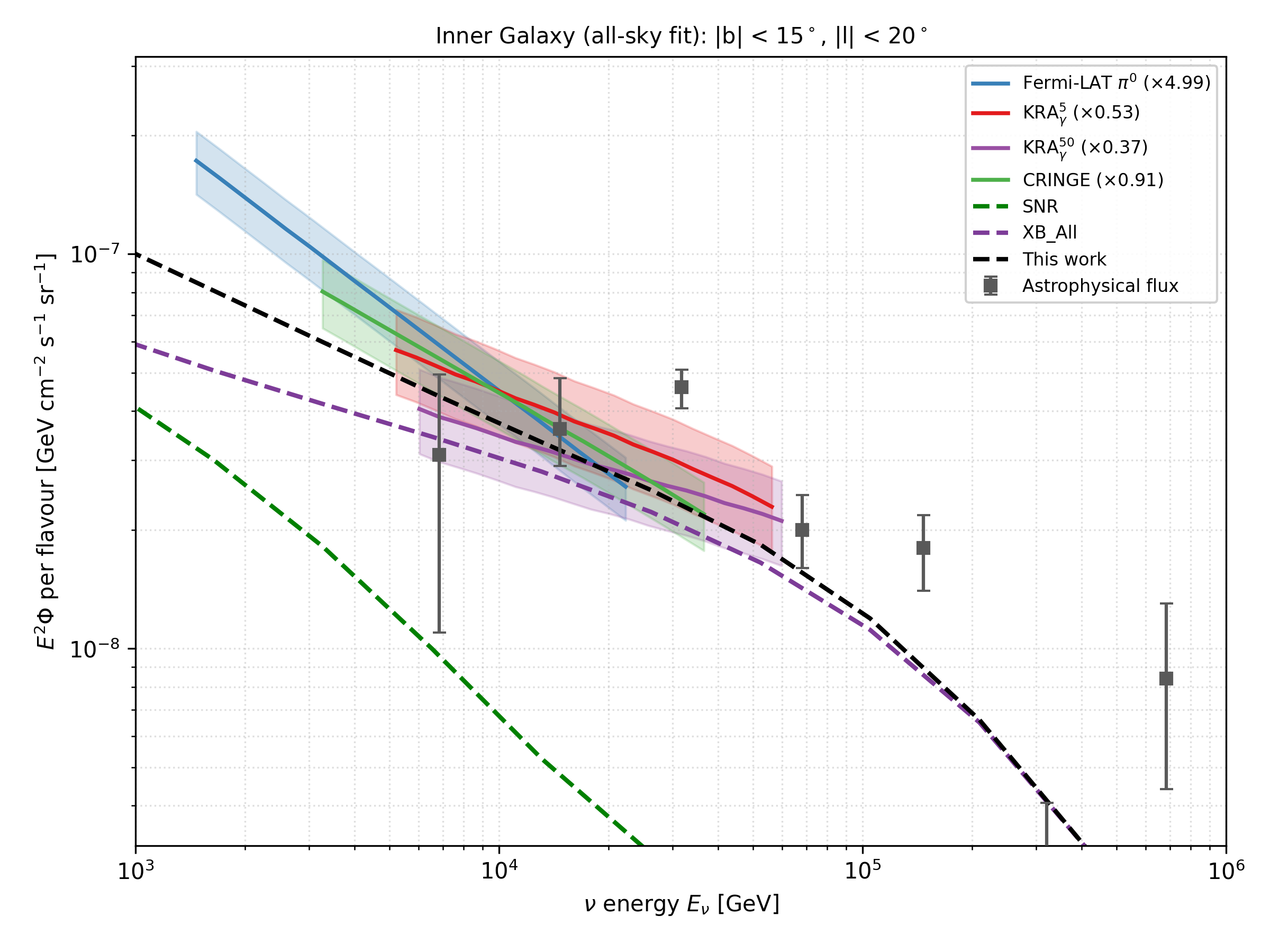}
\includegraphics[width=\columnwidth]{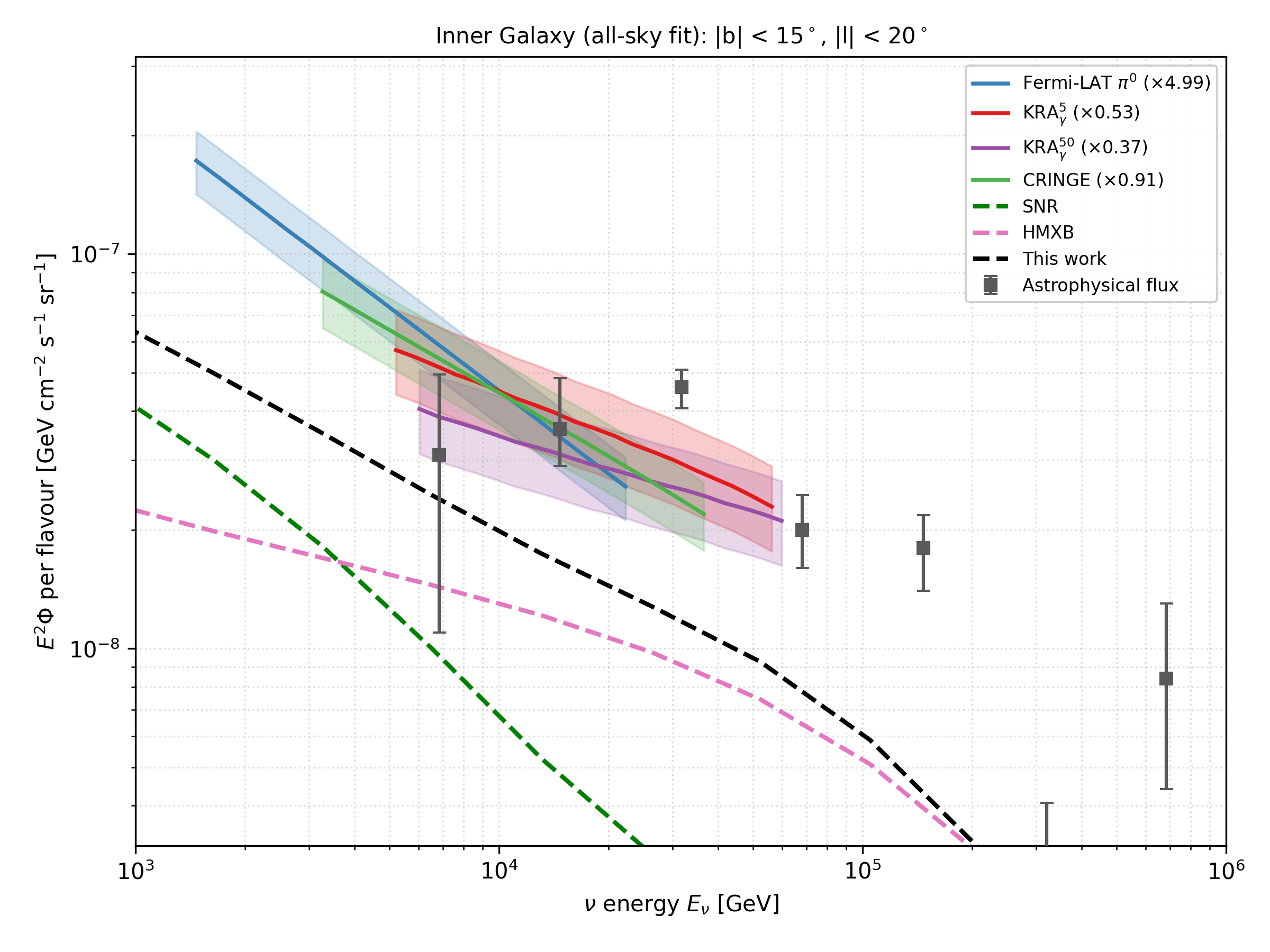}
\includegraphics[width=\columnwidth]{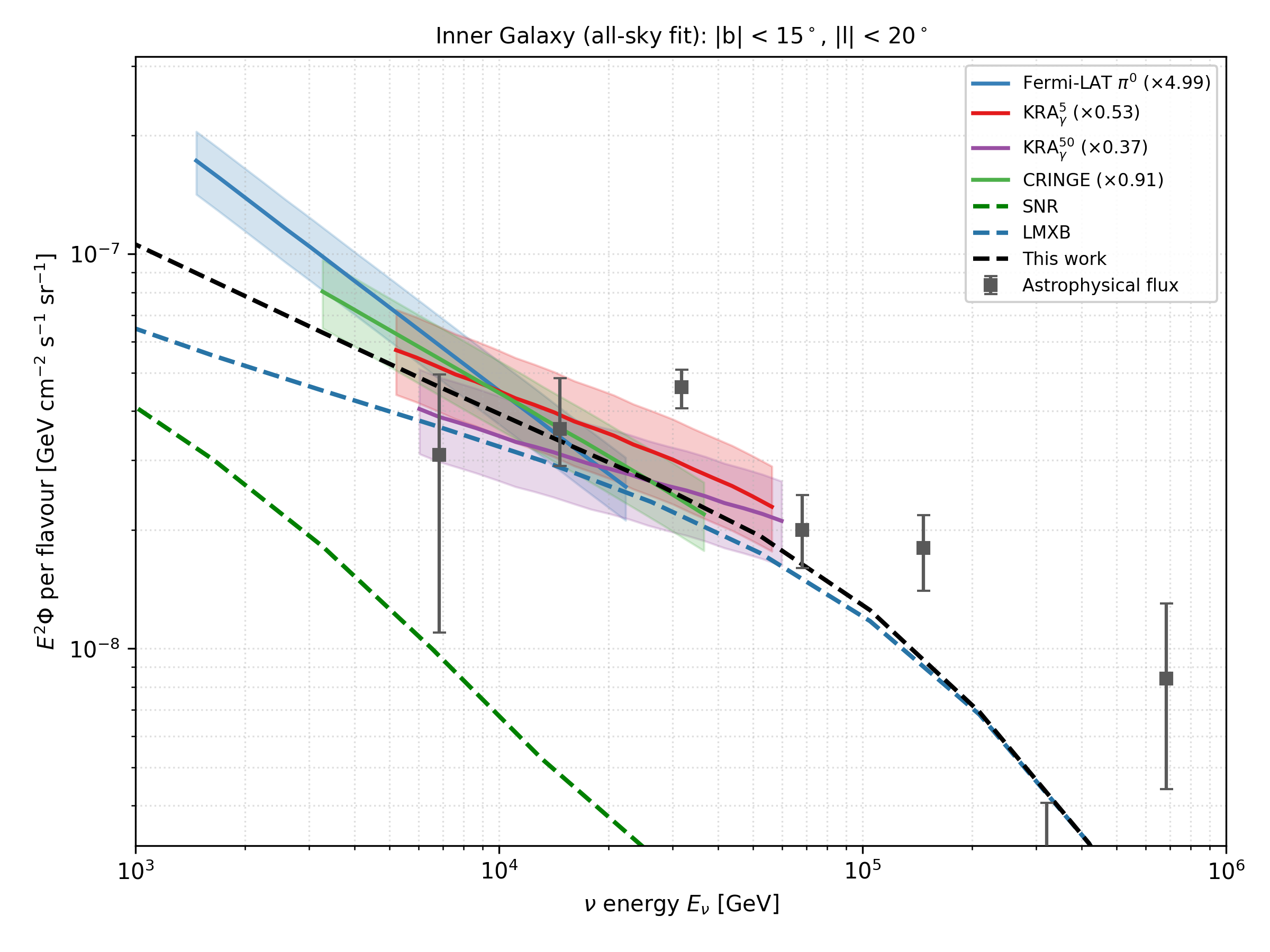}

\caption{Predicted per-flavor diffuse neutrino flux from the inner Galaxy ($|b|<15^{\circ}$, $|l|<20^{\circ}$) for the XB\_All (top), HMXB-dominated (middle), and LMXB-dominated (bottom) scenarios. In each panel, the green dashed lines show the SNR and XRB components and the black dashed line their sum.
The colored bands show the Galactic diffuse neutrino flux in the same region inferred by IceCube\citep{2026arXiv260725966A}, represented by the Fermi-LAT $\pi^{0}$, KRA$_{\gamma}^{5}$,
KRA$_{\gamma}^{50}$, and CRINGE templates scaled to their best-fit normalizations indicated in the legend.}
\label{fig_neutrino}
\end{figure}

The same hadronic interactions that give rise to the $\pi^{0}$-decay $\gamma$-rays inevitably produce neutrinos. When the CR protons and nuclei accelerated by SNRs and XRBs collide inelastically with the interstellar gas, neutral and charged pions are produced with an average number ratio of about $1:2$. The neutral pions decay into pairs of $\gamma$-ray photons, while the charged pions produce neutrinos through the decay chain $\pi^{\pm}\to\mu^{\pm}+\nu_{\mu}(\bar{\nu}_{\mu})$ followed by $\mu^{\pm}\to e^{\pm}+\nu_{e}(\bar{\nu}_{e})+\bar{\nu}_{\mu}(\nu_{\mu})$, i.e., two muon-flavor and one electron-flavor neutrinos per charged pion. The flavor ratio at production is therefore $(\nu_{e}:\nu_{\mu}:\nu_{\tau})\simeq(1:2:0)$, which is reshaped by flavor oscillations during propagation into nearly $(1:1:1)$ by the time the neutrinos arrive at the Earth.

Since the $\gamma$-rays and the neutrinos originate from the decay of the same parent pions, their fluxes obey a fixed conversion relation. Each $\pi^{0}$ produces two photons, and each $\pi^{\pm}$ produces three neutrinos --- one per flavor after flavor oscillations --- so that equal numbers of photons and per-flavor neutrinos are generated per collision. Meanwhile, each secondary particle carries approximately a fixed fraction of the parent pion energy, $E_{\gamma}\simeq E_{\pi}/2$ and $E_{\nu}\simeq E_{\pi}/4$, namely $E_{\gamma}\simeq2E_{\nu}$. The per-flavor neutrino flux can thus be derived directly from the $\pi^{0}$-decay $\gamma$-ray flux as
\begin{equation}
  \phi_{\nu_{\alpha}}(E_{\nu}) \simeq 2\,\phi_{\gamma}(2E_{\nu}), \qquad \alpha=e,\mu,\tau,
  \label{eq:gamma2nu}
\end{equation}
where the factor of two arises from the energy relation $E_{\gamma}\simeq2E_{\nu}$ between the two kinds of secondaries. We emphasize that only the hadronic ($\pi^{0}$-decay) component of the $\gamma$-ray emission enters Eq.~(\ref{eq:gamma2nu}): bremsstrahlung and ICS produce $\gamma$-rays but no neutrinos. Diffuse neutrinos therefore constitute a pure probe of the hadronic interactions of Galactic CRs.

Figure~\ref{fig_neutrino} shows the predicted per-flavor flux from the inner Galaxy ($|b|<15^{\circ}$, $|l|<20^{\circ}$) for the three XB scenarios, in comparison with the Galactic diffuse neutrino flux recently established by IceCube at $5.7\sigma$ in the same region \citep{2026arXiv260725966A}. Following \citep{2026arXiv260725966A}, the measured flux level is represented by the four conventional Galactic emission templates --- the Fermi-LAT $\pi^{0}$ template, KRA$_{\gamma}^{5}$, KRA$_{\gamma}^{50}$, and CRINGE --- scaled to their best-fit normalizations of $4.99^{+0.90}_{-0.86}$,
$0.53^{+0.14}_{-0.12}$, $0.37^{+0.10}_{-0.09}$, and $0.91\pm0.18$, respectively; the spread of the four post-fit spectra reflects the current uncertainty of the measurement. The all-sky-averaged astrophysical neutrino intensity measured by IceCube \citep{2026PhRvL.136l1002A} is also shown for reference. Note that this intensity is dominated by extragalactic sources --- the Galactic component accounts for only $\sim 10\%$ of it at $E_{\nu}\sim 20$~TeV \citep{2026arXiv260725966A} --- so it serves as a reference rather than a benchmark for the Galactic emission.

Similar to the $\gamma$-ray case, the SNR component dominates below a few TeV and then cuts off rapidly, while the XB contribution takes over and dominates the flux at $\gtrsim 10$~TeV. The total fluxes predicted in the XB\_All and LMXB-dominated scenarios fall within the spread of the post-fit IceCube spectra over the sensitive
energy range; notably, unlike the conventional templates, which have to be rescaled by factors of $0.37$--$4.99$, our predictions require no ad-hoc normalization, since the model has already been fixed by the local CR spectra and the LHAASO $\gamma$-ray
data. The HMXB-dominated scenario, on the other hand, predicts a flux about a factor of two lower than the other two scenarios and lies below the lower envelope of the post-fit IceCube spectra at $\sim 10$--$100$~TeV, and is therefore disfavored by the IceCube measurement. Upcoming measurements of the diffuse neutrino flux from the Galactic plane with IceCube and KM3NeT will further test these predictions and may discriminate between the remaining XB scenarios.
\section{Summary}
\label{sec:summary}

The maximum energy that Galactic cosmic rays (CRs) can reach, and the sources responsible for it, remain open questions in high-energy astrophysics. A growing body of very-high-energy gamma-ray observations has revealed the presence of PeV CRs in the Galaxy. However, their sources remain unknown. There is a lack of theoretical and observational evidence demonstrating that SNRs can accelerate CRs up to PeV energies and thus be the primary PeVatron candidates.

The LHAASO and other experiments found a dozen of candidate PeV sources. Among them, the Galactic microquasars have attracted particular attention. A microquasar is an X-ray binary (XRB) system that launches and collimates relativistic jets. However, due to the sparsity of detected microquasars in Galaxy, we do not know their actual spatial distribution. In fact, the XRBs can be categorized into two classes, depending on the mass of the optical companion:  one is LMXBs and the other is HMXBs. The microquasars can belong to either of them.

By the aid of the latest catalog LMXBs and HMXBs, we construct the spatial distribution separately. We find that the diffuse gamma-ray distributions along the Galactic longitude are quite different due to the spatial distributions of these two kinds of sources. When the very high energy gamma rays are dominated by LMXBs, the gamma-ray distribution would be more concentrated toward the direction of the Galactic center. Meantime, we also calculate the neutrino spectrum and find that the HMXBs distribution cannot account for the latest observations of Galactic diffuse neutrinos. 

More precise search of ultrahigh-energy gamma rays and neutrinos could help distinguish the spatial distribution of microquasars can be observed by LHAASO and HAWC in the future, and will confirm the origin of microquasars to high-energy cosmic rays.

\begin{acknowledgments}
We thank Hua Yue for helpful discussions and valuable suggestions. This work is supported by the National Key R\&D Program of China (Grant No. 2025SKA0110103), the National Natural Science Foundation of China (Grants No. 12333006, 12275279, 12575120), the Open Research Fund of the Key Laboratory of Cosmic Rays (Xizang University), Ministry of Education(Grant No. 00061411/008/KLCR2026005), and the Doctoral Innovation Project of Tangshan University (No. 2025065ZB).
\end{acknowledgments}

\bibliographystyle{unsrt_update}
\bibliography{ref}

\end{document}